\documentclass[prb,reprint]{revtex4-1}

\usepackage{graphicx}
\usepackage{bm}
\usepackage{hyperref}
\usepackage{color}
\begin{document}

\title{Multicomponent fractional quantum Hall states with\\ subband
  and spin degrees of freedom}

\date{today}

\author{Yang Liu}
\affiliation{Department of Electrical Engineering,
Princeton University, Princeton, New Jersey 08544}
\author{S.\ Hasdemir}
\affiliation{Department of Electrical Engineering,
Princeton University, Princeton, New Jersey 08544}
\author{J.\ Shabani}
\affiliation{Department of Electrical Engineering,
Princeton University, Princeton, New Jersey 08544}
\author{M.\ Shayegan}
\affiliation{Department of Electrical Engineering,
Princeton University, Princeton, New Jersey 08544}
\author{L.N.\ Pfeiffer}
\affiliation{Department of Electrical Engineering,
Princeton University, Princeton, New Jersey 08544}
\author{K.W.\ West}
\affiliation{Department of Electrical Engineering,
Princeton University, Princeton, New Jersey 08544}
\author{K.W.\ Baldwin}
\affiliation{Department of Electrical Engineering,
Princeton University, Princeton, New Jersey 08544}

\date{\today}

\begin{abstract}
  In two-dimensional electron systems confined to wide GaAs quantum
  wells we observe a remarkable sequence of transitions of the
  fractional quantum Hall states in the filling factor range $1<\nu<3$
  near the crossing of two $N=0$ Landau levels from different subbands
  and with opposite spins. The transitions attest to the interplay
  between the spin and subband degrees of freedom and can be explained
  as pseudospin polarization transitions where the pseudospins are the
  two crossing levels. The magnetic field positions of the transitions
  yield a new and quantitative measure of the transition energies or,
  equivalently, the composite Fermions' discrete energy level
  separations. Surprisingly, these energies are much larger than those
  reported for other systems with SU(2) symmetry; moreover, they are
  larger when the electron system is less spin polarized.
\end{abstract}
 
\maketitle

\section{Introduction}
 
Fractional quantum Hall states (FQHSs) are the hallmarks of an
interacting two-dimensional electron system (2DES) at large
perpendicular magnetic field ($B_{\perp}$) when a Landau level (LL) is
partially occupied.\cite{Tsui.PRL.1982, Jain.CF.2007} Adding extra
electronic (pseudospin) degrees of freedom leads to additional sets of
LLs which are separated by, e.g., the Zeeman energy ($E_Z$), subband
separation ($\Delta_{SAS}$), or valley splitting energy ($E_V$), in
systems with spin, subband, or valley degree of freedom,
respectively. When two such LLs cross at a particular LL filling
factor ($\nu$), the QHS at that filling typically weakens or
disappears.\cite{Liu.PRB.2011} In very high quality samples, however,
if two $N=0$ LLs cross, the 2DES can form two-component FQHSs where
the components are the crossing LLs (the pseudospins), and exhibit
pseudospin polarization transitions as one tunes the pseudospin energy
splitting, or the separation between the two LLs. Such transitions
have been reported for 2DESs confined to AlAs quantum wells (QWs)
where the energy separation between the two occupied valleys is tuned
via the application of strain so that the two lowest ($N=0$) LLs of
the two valleys cross.\cite{Bishop.PRL.2007, Padmanabhan.PRB.2009,
  Padmanabhan.PRB.2010, Padmanabhan.PRL.2010} Multi-component FQHSs
and transitions between their pseudospin configurations have also been
studied in other 2DESs where two or more LLs with different spin or
valley indices are close in energy.\cite{Eisenstein.PRL.1989,
  Clark.PRL.1989, Engel.PRB.1992, Du.PRL.1995, Kukushkin.PRL.1999,
  Liu.PRB.2014B, Kott.PRB.2014, Dean.Nat.Phys.2011, Feldman.PRL.2013,
  Falson.Nat.Phys.2015}
 
Here we present a study of 2DESs confined to wide GaAs QWs where $E_Z$
and $\Delta_{SAS}$ are much smaller than the Coulomb energy at large
$B_{\perp}$. Thus the lowest four LLs -- the S0$\uparrow$,
S0$\downarrow$, A0$\uparrow$ and A0$\downarrow$ levels -- are close in
energy so that, in principle, a full description of FQHSs would
require the inclusion of all four LLs and using SU(4) symmetry (S and
A refer to symmetric and antisymmetric subbands, $\uparrow$ and
$\downarrow$ refer to up- and down-spin; see Fig. 1(a)). Via applying
either a parallel magnetic field component ($B_{||}$, see Fig. 1(b)),
or changing the electron density in the QW,\cite{Note1}
we reduce
$\Delta_{SAS}$ and increase $E_Z$, so that the S0$\downarrow$ and
A0$\uparrow$ levels cross when $\Delta_{SAS}=E_Z$; see Fig. 1(a). Near
the crossing, we observe a remarkable pattern of appearing and
disappearing FQHSs in the filling range $1<\nu<3$, revealing the
formation of FQHSs with both spin and subband degrees of freedom
simultaneously.\cite{Note2}
We can qualitatively describe the number
of observed transitions and their positions in a simplified
two-component picture with SU(2) symmetry, where the two pseudospins
are the S0$\downarrow$ and A0$\uparrow$ levels. However, our
quantitative analysis reveals several puzzling features. We find that
at a fixed $\nu$, the transition energies are much larger than
reported previously in systems with SU(2) symmetry where the two
pseudospins are either spin or valley levels. Also, the transition
energies are about twice larger when the 2DES is less
spin-polarized. We discuss possible origins of these unexpected
features.

\begin{figure*}
\includegraphics[width=0.95\textwidth]{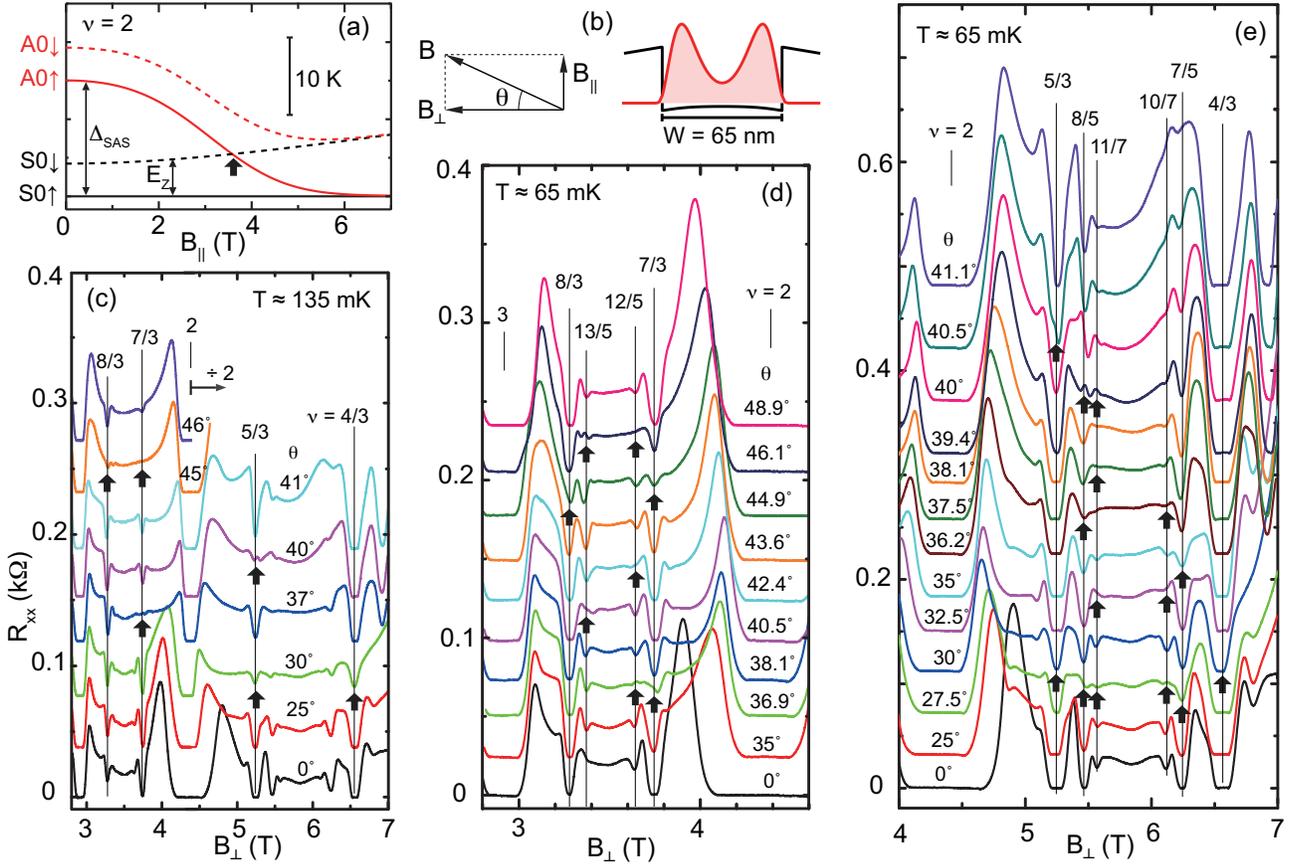}
\caption{\label{fig:waterfall} (color online) (a) Relevant energy levels for
  $n=2.12\times 10^{11}$ cm$^{-2}$ electrons confined to a 65-nm-wide
  GaAs QW near filling factor $\nu=2$, calculated as a function of
  parallel magnetic field ($B_{||}$). (b) Charge distribution (red
  curve) and potential (black curve) calculated via solving
  Poisson-Schr\"odinger equations self-consistently at $B=0$. The
  experimental geometry and the tilt angle $\theta$ are also
  shown. (c)-(e) Waterfall plots of $R_{xx}$ vs $B_{\perp}$, measured
  at the indicated temperatures and different $\theta$. Each trace is
  shifted vertically. The observed FQHS transitions are marked by the
  black arrows.}
\end{figure*}

\section{Method}

Our sample, grown by molecular beam epitaxy, is a 65-nm-wide GaAs QW
bounded on each side by undoped Al$_{0.24}$Ga$_{0.76}$As spacer layers
and Si $\delta$-doped layers. We use a 4 $\times$ 4 mm$^{2}$ square
piece with alloyed InSn contacts at four corners, and an evaporated
Ti/Au front-gate and an In back-gate to change the 2DES density ($n$)
while keeping the charge distribution symmetric [Fig. 1(b)]. We
measure the transport coefficients in a dilution refrigerator with
base temperature $T\approx 30$ mK, using low-frequency ($<$ 30 Hz)
lock-in technique and a rotatable sample platform to induce
$B_{||}$. At $B_{||}=0$, the Fourier transform of the Shubnikov-de
Haas oscillations exhibits two peaks that correspond to the electron
densities in the two subbands. The difference between these densities
yields $\Delta_{SAS}$ which is in excellent agreement with the results
of our $B=0$ self-consistent calculations. To calculate $\Delta_{SAS}$
at finite $B_{||}$ and $B_{\perp}$ [Fig. 1(a)], we employ a
perturbative simulation introduced in Ref. [\onlinecite{Kumada.PRB.2008}],
where we assume $B_{||}$ only mixes LLs from different subbands but
does not change the QW potential.

\section{Experimental results}

Figure 1 (c) shows the longitudinal magnetoresistance ($R_{xx}$)
traces for $1<\nu<3$, measured at $n=2.12\times 10^{11}$ cm$^{-2}$,
$T\simeq 135$ mK, and different tilt angles ($\theta$). As we increase
$\theta$, the $\nu=2$ $R_{xx}$ plateau narrows near
$\theta\simeq 37^{\circ}$ and then widens again at larger
$\theta$. The reduction of the $R_{xx}$ plateau width signals a
weakening of the integer QHS, which is widely used to identify LL
crossings in high quality 2DESs and at low temperatures.\cite{Liu.PRB.2011, Desrat.PRB.2005, Zhang.PRB.2006} This weakening,
but not disappearing, is similar to what is seen at certain LL
crossings at other integral $\nu$ when interaction preserves the
energy gap through QHS ferromagnetism.\cite{Poortere.Science.2000,
  Muraki.PRL.2001, Liu.PRB.2011} In our samples, $\Delta_{SAS}$ is
larger than $E_Z$ at $B_{||}=0$. As we increase $B_{||}$,
$\Delta_{SAS}$ decreases and $E_Z$ increases, so that the
S0$\downarrow$ and A0$\uparrow$ levels cross when
$\Delta_{SAS}\simeq E_Z$.\cite{Note3}
In Fig. 1(a), we show the calculated energies of the four $N=0$ LLs
relative to the S0$\uparrow$ level as a function of $B_{||}$ at
$\nu=2$.\cite{Kumada.PRB.2008} We use a three-fold enhanced $E_Z$ to
match the experimental observation that the $\nu=2$ crossing occurs
near $\theta\simeq 37^{\circ}$ ($B_{||}\simeq 3.5$ T). This
enhancement of $E_Z$ is not surprising and has also been reported in
previous studies.\cite{Liu.PRB.2011, Falson.Nat.Phys.2015,
  Desrat.PRB.2005, Zhang.PRB.2006}

More interestingly, near $\theta\simeq37^{\circ}$, the FQHSs on both
sides of $\nu=2$ show a rich series of transitions, as marked with
arrows in Figs. 1(c)-(e).\cite{Note4}
The $\nu=4/3$ FQHS is strong at both small and large $\theta$, but
becomes weak at $\theta\simeq 30^{\circ}$ [Figs. 1(c) and (e)]. The
$\nu=8/3$ FQHS also experiences one transition, at
$\theta\simeq 45^{\circ}$ [Figs. 1(c) and (d)]. Meanwhile, the
$\nu=5/3$ and 7/3 FQHSs become weak twice: at
$\theta\simeq 30^{\circ}$ and 40$^{\circ}$ for $\nu=5/3$, and at
$\theta\simeq 37^{\circ}$ and 45$^{\circ}$ for $\nu=7/3$. Data taken
at lower $T\simeq 65$ mK, shown in Figs. 1(d) and (e), reveal a more
remarkable pattern of higher-order FQHS transitions. On the left side
of $\nu=2$ [Fig. 1(d)], the $\nu=13/5$ FQHS weakens twice at
$\theta\simeq 40.5^{\circ}$ and $46.1^{\circ}$. The $\nu=12/5$ FQHS,
on the other hand, becomes weak three times, at
$\theta\simeq36.9^{\circ}$, $42.4^{\circ}$ and $46.1^{\circ}$. On the
high-field side of $\nu=2$, as seen in Fig. 1(e), the $\nu=7/5$ FQHS
weakens twice, at $\theta\simeq 27.5^{\circ}$ and $35^{\circ}$, and
the $\nu=8/5$ FQHS thrice, at $\theta\simeq27.5^{\circ}$,
$36.2^{\circ}$ and $39.4^{\circ}$. We also observe in Fig. 2(e) three
transitions at $\nu=10/7$ and four transitions at $\nu=11/7$.\cite{Note1} We summarize in Fig. 2(a) the values of $B_{||}$ and
$B_{\perp}$ for all the observed transitions.

\begin{figure}
\includegraphics[width=.45\textwidth]{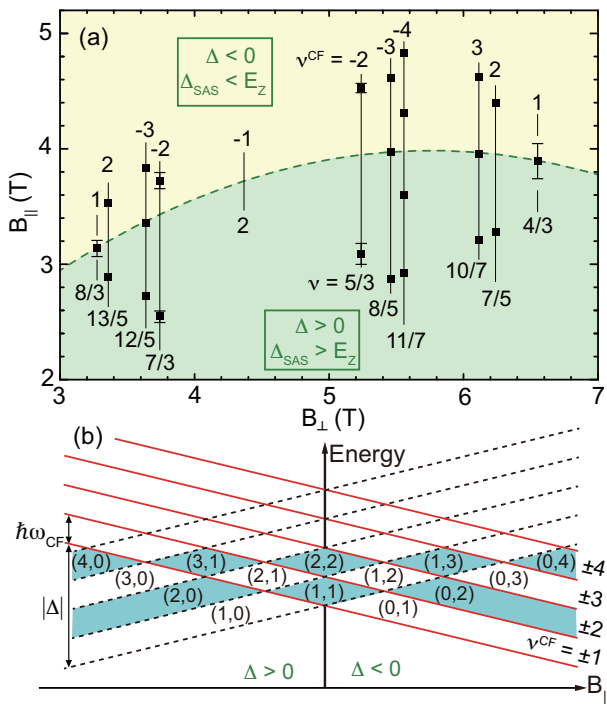}
\caption{\label{fig:cartoon}(color online) (a) Summary of $B_{||}$ and
  $B_{\perp}$ where pseudospin transitions are seen in Fig. 1. The
  dashed line represenst zero pseudospin splitting $\Delta=0$ (see
  text). $\Delta=\Delta_{SAS}-E_Z$ is positive in the green region,
  and becomes negative in the yellow region. Typical error bars are
  shown for the $q/3$ FQHS transitions. (b) Schematic depiction of the
  CF $\Lambda$-level energies in a simplified two-pseudospin
  picture. The dashed black and solid red lines represent the CF
  $\Lambda$-levels with up- and down-pseudospins (the electron
  S0$\downarrow$ and A0$\uparrow$ levels). Each configuration is
  labeled with its pseudospin $\Lambda$-level filling factors
  ($\nu_{\text{CF}\uparrow}$, $\nu_{\text{CF}\downarrow}$), where
  $\uparrow$ and $\downarrow$ stand for the up- and down-pseudospins.}
\end{figure}

\section{Discussion}

The fact that all these transtions occur near the crossing of the
S0$\downarrow$ and A0$\uparrow$ levels when $\Delta_{SAS}\simeq E_Z$
suggests that these are FQHS pseudospin polarization transitions.\cite{Note2} Such transitions can be readily understood in the
framework of composite Fermions (CFs), quasiparticles formed by
attaching two flux quanta to each electron.\cite{Jain.PRL.1989,
  Halperin.PRB.1993, Jain.CF.2007} The CFs form their own discrete
energy levels, the so-called $\Lambda$-levels which are separated by
the CF cyclotron energy $\hbar\omega_{\text{CF}}$, and the FQHS at
$\nu=\nu_{\text{CF}}/(2\nu_{\text{CF}}+1)$ is the integer QHS of CFs
with integral $\Lambda$-level filling factor $\nu_{\text{CF}}$. Since
the other two energy levels, S0$\uparrow$ and A0$\downarrow$, are
reasonably far in energy and are always full or empty,\cite{Note2} 
we neglect them and interpret our data in a simple (SU(2)) picture
where the two crossing levels, S0$\downarrow$ and A0$\uparrow$, are
the two pseudospins.\cite{Note5} 
The pseudospin splitting energy
$\Delta=\Delta_{SAS}-E_Z$ is then tuned from positive to negative via
applying $B_{||}$ [Figs. 1(a) and 2(b)]. As discussed below, this
simplified two-pseudospin model qualitatively explains our data.

In this model, the FQHS at $\nu=4/3$ has only one occupied $\Lambda$
level ($\nu_{\text{CF}}=1$).\cite{Note5} It therefore has two
possible configurations with the $\Lambda$-level filling factors of
the up- and down-pseudospin CFs being
$(\nu_{\text{CF}\uparrow},\nu_{\text{CF}\downarrow})=(1,0)$ or (0,1),
and should exhibit one transition at $\Delta=0$ [see Fig. 2(b)],
consistent with Fig. 2(a) data. The FQHS at $\nu=7/5$, on the other
hand, has two filled $\Lambda$-levels ($\nu_{\text{CF}}=2$), three
pseudospin configurations
[$(\nu_{\text{CF}\downarrow}, \nu_{\text{CF}\downarrow})=$ (2,0),
(1,1) and (0,2)], and two transitions when
$\Delta=\pm\hbar\omega_{\text{CF}}$ [Fig. 2(b)], also in agreement
with Fig. 2(a) data. Using similar logic, we can explain the three
transitions observed for the $\nu=10/7$ FQHS which has
$\nu_{\text{CF}}=3$ and four configurations [Fig. 2(b)]. By invoking
particle-hole symmetry, which links the FQHSs at $\nu$ to $(4-\nu)$ in
our system (e.g., 4/3 to 8/3), and also utilizing negative CF
fillings, e.g., $\nu_{\text{CF}}=-2$ for the $\nu=5/3$ state, we can
explain \emph{all} the transitions summarized in Fig. 2(a). In
general, a FQHS with $\nu_{\text{CF}}$ has $|\nu_{\text{CF}}|+1$
pseudospin configurations, and $|\nu_{\text{CF}}|$ pseudospin
polarization transitions which occur whenever $|\Delta|$ equals a
multiple integer of $\hbar\omega_{\text{CF}}$ [see Fig. 2(b)].

Next we proceed to a quantitative analysis of Fig. 2(a) data.  Figure
2(b) implies that, when $\nu_{\text{CF}}$ is odd, one transition
occurs exactly at $\Delta=0$. In Fig. 2(a), we first fit the dashed
line through these transition points. We then focus on a
particular $\nu$, e.g. 8/5, and calculate $\Delta_{SAS}$ as a function
of $B_{||}$ at this filling, as shown in Fig. 3(a). Using the value of
$B_{||}$ at which the $\Delta=0$ transition for $\nu=8/5$ occurs
($B_{||}=3.96$ T), we determine a value for $g^*$ ($\simeq -1.34$) so
that $\Delta_{SAS}=E_Z$ at $B_{||}=3.96$ T. We then plot $E_Z$
($=g^*\mu_B B$) at $\nu=8/5$ as a function of $B_{||}$ in Fig. 3(a),
and determine $\Delta$ for the other two 8/5 transitions. (See
Fig. 3(b) for the four pseudospin configurations of the $\nu=8/5$
FQHS.) Since these transitions are expected to occur when
$\Delta=\pm2\hbar\omega_{\text{CF}}$ [see Fig. 2(b)], we find
$\hbar\omega_{\text{CF}}=5.5$ K for $\Delta>0$ and
$\hbar\omega_{\text{CF}}=2.5$ K for $\Delta<0$. Using this procedure
we can deduce $\hbar\omega_{\text{CF}}$ for FQHSs at $\nu=$ 8/5, 12/5,
and 10/7 which have a $\Delta=0$ transition on the dashed line in
Fig. 3(a). For the FQHSs at $\nu=$ 13/5, 7/3, 5/3 and 7/5, we assume
$\Delta=0$ at the intersection of the dashed line and the vertical lines
that mark these fillings in Fig. 3(a) and, following a similar
procedure, find $\hbar\omega_{\text{CF}}$ from the transitions'
$B_{||}$ values.

\begin{figure}
\includegraphics[width=.45\textwidth]{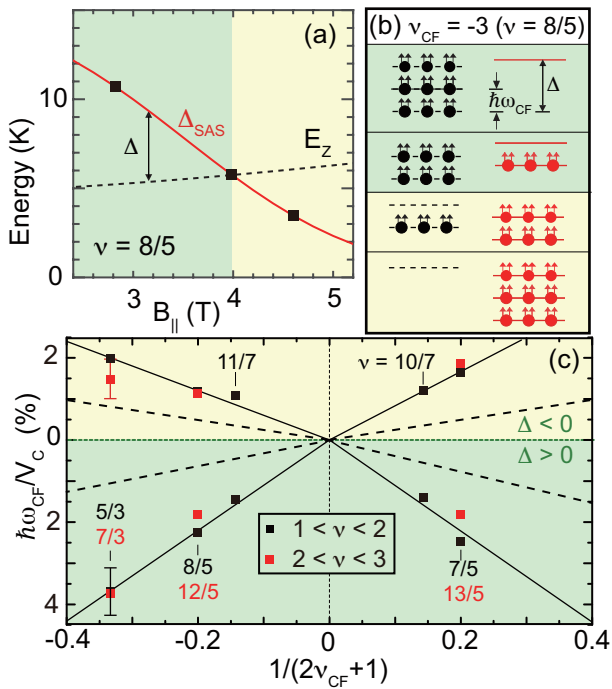}
\caption{\label{fig:cartoon}(color online) (a) Calculated
  $\Delta_{SAS}$ and $E_Z$ as a function of $B_{||}$ at $\nu=8/5$. (b)
  $\Lambda$-level diagram showing the four different pseudospin
  configurations for $\nu=8/5$ ($\nu_{\text{CF}}= -3$) in our
  two-component picture. The pseudospin polarization transitions are
  expected when $\Delta=0, \pm2\hbar\omega_{\text{CF}}$ [see
  Fig. 2(b)]. (c) The $\Lambda$-level separation in units of
  $V_C=e^2/(4\pi\epsilon l_B)$ as a function of
  $1/(2\nu_{\text{CF}}+1)$, deduced from $\Delta$ at which the FQHS
  transitions are observed in Fig. 2(a). Data are shown for both
  $1<\nu<2$ (black) and $2<\nu<3$ (red). The dashed lines represent
  data for 2DESs with only spin degree of freedom
  \cite{Liu.PRB.2014B}; results from transitions near $\nu=1/2$ and 3/2
  are shown in the $\Delta<0$ and $\Delta>0$ regions, respectively.}
\end{figure}

In Fig. 3(c) we plot as a function of $1/(2\nu_{\text{CF}}+1)$ all the
deduced values of $\hbar\omega_{\text{CF}}$, normalized to the Coulomb
energy ($V_C=e^2/4\pi\epsilon l_B$, where
$l_B=\sqrt{\hbar/eB_{\perp}}$ is the magnetic length and $\epsilon$ is
the dielectric constant), for different FQHSs. Figure 3(c) data allow
us to make a quantitative comparison of the deduced
$\hbar\omega_{\text{CF}}/V_C$ to previous experimental reports. The
dashed lines in Fig. 3(c) represent $\hbar\omega_{\text{CF}}/V_C$,
measured near $\nu=1/2$ and 3/2, from FQHS \emph{spin} transitions in
QWs with same well-width (65 nm) but much lower density where the
S0$\uparrow$ and S0$\downarrow$ levels are closer in energy and the
A0$\uparrow$ level is well above S0$\downarrow$.\cite{Liu.PRB.2014B, Note1, Note2} Figure 3(c) plot clearly shows that our
measured $\hbar\omega_{\text{CF}}/V_C$ is much larger than the dashed
lines. 

It is possible that the transition energies we deduce are
somewhat exaggerated because of the inaccuracy of the perturbative
calculations we use to determine the $B_{||}$-dependence of
$\Delta_{SAS}$. To test this possibility, we made measurements on
another QW with slightly narrower well width (55 nm).\cite{Note1} In
this sample, we induce the crossing of the S0$\downarrow$ and
A0$\uparrow$ levels via applying $B_{||}$ and also by tuning $n$ at
$B_{||}=0$. The $\hbar\omega_{\text{CF}}$ deduced from the tilting
data matches Fig. 3 data, but the $n$-tuning results are somewhat
smaller,\cite{Note1} suggesting the inaccuracy of our perturbative
calculations.\cite{Note6} 
However, we emphasize that even the
$n$-tuning data which do not rely on calculations of $\Delta_{SAS}$,
yield $\hbar\omega_{CF}/V_C$ values that are about a factor of two
larger than the dashed lines in Fig. 3(c).\cite{Note1} We therefore
conclude that $\hbar\omega_{\text{CF}}$ in our 2DES with both spin and
subband degrees of freedom is larger than in a 2DES with only spin
degree of freedom. The reason for this is not entirely clear. Perhaps
the nature of pseudospins plays a role in determining the polarization
energies and $\hbar\omega_{CF}$. It is also possible that in our
system the close proximity of the other two LLs (S0$\uparrow$ and
A0$\downarrow$; see Fig. 1(a)) is important. Both of these interesting
possibilities deserve further theoretical examination.

We highlight two additional noteworthy features of Fig. 3(c). First,
the results for $1<\nu<2$ approximately match those for $2<\nu<3$,
suggesting that $\hbar\omega_{\text{CF}}$ does not depend on which LL
(S0$\downarrow$ or A0$\uparrow$) hosts the CFs. Second, in our 2DES we
tune the crossing of two $N=0$ LLs with opposite spins while a lower
level (S0$\uparrow$) is fully occupied. This allows us to observe FQHS
pseudospin transitions at a given $\nu$ for both $\Delta<0$ and
$\Delta>0$, when the net spin orientation of the CFs is aligned or
anti-aligned with the spin ($\uparrow$) of majority electrons,
respectively. Data of Fig. 3(c) indicate that
$\hbar\omega_{\text{CF}}/V_C$ is about twice larger for the case
$\Delta>0$ compared to the $\Delta<0$ case. This is an important
observation as it sheds light on the mysterious asymmetry between
$\hbar\omega_{\text{CF}}/V_C$ for CFs near $\nu=3/2$ and 1/2, deduced
from CF spin-polarization transitions in GaAs 2DESs,\cite{Liu.PRB.2014B} or CF valley-polarization transitions in AlAs
2DESs.\cite{Padmanabhan.PRB.2010} In those 2DESs the polarization
transition energies of FQHSs near 3/2, where electrons are partially
spin- (or valley-) polarized, are much larger than those near 1/2 where
they are fully polarized.\cite{Liu.PRB.2014B,
  Padmanabhan.PRB.2010} Our data, afforded by the tunability of
the subband/spin degrees of freedom, provide clear evidence that the
spin polarization of the total electron system can affect the CF
properties.

\begin{acknowledgments}
  We acknowledge support through the NSF (DMR-1305691), and DOE BES
  (DE-FG02-00-ER45841) for measurements, and the NSF (Grant MRSEC
  DMR-1420541), the Gordon and Betty Moore Foundation (Grant
  GBMF4420), and Keck Foundation for sample fabrication and
  characterization. A portion of this work was performed at the NHMFL,
  which is supported by the NSF Cooperative Agreement No. DMR-1157490,
  the State of Florida, and the DOE. We thank S. Hannahs, G. E. Jones,
  T. P. Murphy, E. Palm, A. Suslov, and J. H. Park for technical
  assistance, and J. K. Jain for illuminating discussions.
\end{acknowledgments}

\bibliography{../bib_full}

\end{document}